\documentclass[iop]{emulateapj}
\usepackage{amsmath}
\usepackage{graphicx}
\usepackage{color}

\begin{document}
\title{Ice Lines in Circumbinary Protoplanetary Disks}  

\author{Christian Clanton}
\affil{Department of Astronomy, The Ohio State University, 140 W. 18th Ave., Columbus, OH 43210, USA}
\email{clanton@astronomy.ohio-state.edu}

\begin{abstract}
    I examine the position of the ice line in circumbinary disks heated by steady mass accretion and stellar irradiation and compare with the critical semi-major axis, interior to which planetary orbits are unstable. There is a critical binary separation, dependent on the binary parameters and disk properties, for which the ice line lies within the critical semi-major axis for a given binary system. For an equal mass binary comprised of $1~{\rm M_{\odot}}$ components, this critical separation is $\approx 1.04~$AU, and scales weakly with mass accretion rate and Rosseland mean opacity ($\propto [\dot{M}\kappa_{\rm R}]^{2/9}$). Assuming a steady mass accretion rate of $\dot{M} \sim 10^{-8}~{\rm M_{\odot}~yr^{-1}}$ and a Rosseland mean opacity of $\kappa_{\rm R}\sim 1~{\rm cm^2~g^{-1}}$, I show that $\gtrsim 80\%$ of all binary systems with total masses $M_{\rm tot} \lesssim 4.0~{\rm M_{\odot}}$ have ice lines that lie interior to the critical semi-major axis. This suggests that rocky planets should not form in these systems, a prediction which can be tested by looking for planets around binaries with separations larger than the critical separation with \emph{Kepler} (difficult) and with microlensing.
\end{abstract}


\section{Introduction}
\label{sec:intro}
The ``ice line,'' or ``snow line,'' is the point in protoplanetary disks at (and beyond) which the temperature is low enough that water-ice can condense from the nebular gas. Stellar irradiation, mass accretion, and general disk properties (e.g. opacity, surface mass density distribution) determine the position of the ice line. The position of the ice line is weakly dependent on the disk environment; stellar irradiation, mass accretion, and general disk properties determine the position. The temperature threshold below which ice can form is $\sim 145-170~$K, depending on the partial pressure of the nebular water vapor \citep{2004M&PS...39.1859P,2006ApJ...640.1115L}. The ice line is important in the context of planetary formation because it separates the inner region of rocky planet formation from the outer region of gaseous and icy planet formation \citep{2005ApJ...626.1045I,2008ApJ...673..502K}.

Gas giant formation by core accretion occurs when protoplanet core masses are sufficient to attract gas from the nebula. \citet{2000ApJ...537.1013I} show that the timescale for gas giant formation is set by the core mass \citep[see also][]{2005Icar..179..415H}. Thus, if the core mass is too small, it will not be able to accrete enough gas to form a gas giant before the gas is removed. Past the ice line, condensation of water ices increase the surface density of the disk by a factor of $\sim 3$, increasing the isolation (core) mass by a factor of $\sim 5$ \citep{2008ApJ...673..502K}. This increase in surface density beyond the ice line allows for gas giant formation by core accretion within the short lifetime of disk gas, believed to be somewhere in the range of $\sim$1--10~Myr \citep{1995Natur.373..494Z,2006ApJ...651.1177P}. It is therefore important to know where the snow line lies in protodisks, as it holds much information about where gas giants are formed and how their orbits evolve in time.

Only recently have circumbinary planets been discovered. There are currently six known circumbinary planets in five different systems found by \emph{Kepler} \citep{2011Sci...333.1602D,2012Natur.481..475W,2012ApJ...758...87O,2012Sci...337.1511O}. Additionally, there are 12 eclipsing, close-compact binary systems with white dwarf or hot sdB primaries showing apparent period variations that could perhaps be explained by circumbinary giant planets. These suspected giant planets were probably not formed primordially, but rather it is more likely they formed later, after common envelope evolution (see \citet{2013A&A...549A..95Z} for a detailed discussion of these systems). Nevertheless, the unambiguous discovery of circumbinary planets by \emph{Kepler} has motivated a growing body of literature regarding planetary formation in circumbinary disks, including this study.

In this paper, I study ice line positions in circumbinary disks and investigate the parameters that determine their locations. In \S~\ref{sec:model} I describe a simple circumbinary disk model, and in \S~\ref{sec:results} I calculate ice line positions and describe results. Conclusions are presented in \S~\ref{sec:conclusion}.

\section{Model}
\label{sec:model}

\subsection{Disk Model and Stellar Parameters}
\label{subsec:disk_model}
Similar to the treatment of \citet{2006ApJ...640.1115L} in the case of a single star, I consider a circumbinary disk heated by steady mass accretion as well as stellar irradiation from the binary. I adopt a circumbinary disk with a flared geometry, as described in \citet{1997ApJ...490..368C}. The surface mass density distribution is $\Sigma = \Sigma_0\left(a/{\rm AU}\right)^{-3/2}$, with $\Sigma_0=10^3~$g~cm$^{-2}$ \citep{1977Ap&SS..51..153W}.

Several simulations have shown that the inner edge of circumbinary disks should be truncated between $1.8~a_{\rm bin}$ and $2.6~a_{\rm bin}$, where $a_{\rm bin}$ is the binary semi-major axis, due to tidal torques exerted by the binary \citep{1991ApJ...370L..35A,1994ApJ...421..651A,2002A&A...387..550G,2007A&A...472..993P,2008ApJ...672...83M}. There has been general agreement with these predictions from observations of directly imaged circumbinary disks, most notably that found around GG Tauri by \citet{1994A&A...286..149D} \citep[see also][]{2005A&A...439..585B,1998A&A...332..867D, 2008ApJ...678L..59I}. Furthermore, \citet{1999AJ....117..621H} have shown that there is a critical semi-major axis for planets orbiting a binary. Within this critical semi-major axis, the planets are too strongly affected by the binary and their orbits are unstable. \citet{1999AJ....117..621H} provide the relation
\begin{align}
    a_c = & {} \left[\left(1.60 \pm 0.04\right) + \left(5.10 \pm 0.05\right)e\right. \nonumber \\
    & {} + \left(-2.22 \pm 0.11\right)e^2 + \left(4.12 \pm 0.09\right)\mu \nonumber \\
    & {} + \left(-4.27 \pm 0.17\right)e\mu + \left(-5.09 \pm 0.11\right)\mu^2 \nonumber \\
    & {} + \left.\left(4.61 \pm 0.36\right)e^2\mu^2\right]a_{\rm bin}\; , \label{eqn:a_crit}
\end{align}
where $a_c$ is the critical semi-major axis, $\mu = m_2/\left(m_1+m_2\right)$ is the mass ratio of the binary, and $e$ is the eccentricity of the binary. For an equal mass binary on a circular orbit $\left(\mu = 1/2, e = 0\right)$, this relation yields $a_c \approx 2.4~a_{\rm bin}$, consistent with the truncation radius expected for circumbinary disks.

I construct pre-main sequence binary systems using stellar masses, temperatures, and radii from the models of \citet{2000A&A...358..593S} at an age of one-tenth the stars' zero-age main sequence (ZAMS), $t_{\rm ZAMS}/10$. I examine both equal and unequal mass binaries, with component masses ranging from $0.5-5.0~{\rm M_{\odot}}$.

\subsection{Midplane Temperature Profiles}
\label{subsec:temp_profiles}
{\it Accretion Heating: }As in \citet{2006ApJ...640.1115L}, I adopt the \citet{1974MNRAS.168..603L} prescription for steady accretion
\begin{equation}
    T_{\rm eff, acc}^4 = \frac{3}{8\pi \sigma}\frac{GM_{\star}\dot{M}}{r^3}\; . \label{eqn:t_effacc}
\end{equation}
To generalize this to the case of a binary, I replace $M_{\star}$ with the total mass of the binary, $M_{\rm tot}$. Note that I ignore the $\sqrt{R_{\star}/r}$ term usually associated with equation \ref{eqn:t_effacc}, as it is a very small correction for this study due to the truncated inner edges of circumbinary disks. I adopt a constant value of $\dot{M} = 10^{-8}~{\rm M_{\odot}~yr^{-1}}$ for the mass accretion rate\footnote{\citet{1998ApJ...495..385H} found a median accretion rate for T Tauri stars of $\sim 10^{-8}~{\rm M_{\odot}~yr^{-1}}$, with a spread as large as one order of magnitude, in the Taurus and Chamaeleon I molecular cloud complexes. I discuss the sensitivity of my results to this choice in \S~\ref{sec:results}.}.

The temperature due to accretional flux diffusing in an optically thick medium from the midplane to the surface of the disk is
\begin{equation}
    T_{\rm acc}^4 = \frac{3}{4}\left(\tau_{\rm R}+\frac{2}{3}\right)T_{\rm eff, acc}^4\; , \label{eqn:t_acc}
\end{equation}
where $\tau_{\rm R}$ is the Rosseland optical depth
\begin{equation}
    \tau_{\rm R} = \displaystyle \int_0^\infty \kappa_{\rm R} \rho\left(r,z\right)dz\; ,
\end{equation}
which I approximate as $\tau_{\rm R}\sim \kappa_{\rm R} \Sigma\left(r\right)/2$, where $\kappa_{\rm R}$ is the Rosseland mean opacity. I adopt the opacity value of $\kappa_{\rm R}\approx 1~{\rm cm^2~g^{-1}}$ from \citet{2001ApJ...553..321D}, using their dust model with a maximum grain size of $a_{\rm max}=1~$mm and grain size distribution power law slope of $p=3.5$. \citet{2004ApJ...608..497J} showed that the dependence of this opacity on the temperature structure of a disk is small between $100-300$~K, so I assume that $\kappa_{\rm R}$ is constant throughout the regions of the disk I examine.

\begin{figure*}
 \begin{picture}(0,0)%
\includegraphics{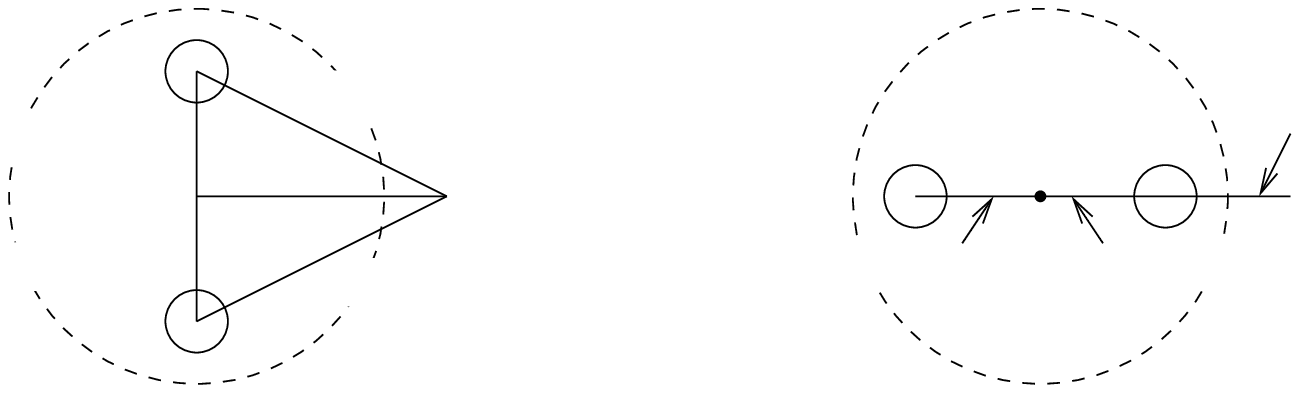}%
\end{picture}%
\setlength{\unitlength}{3947sp}%
\begingroup\makeatletter\ifx\SetFigFont\undefined%
\gdef\SetFigFont#1#2#3#4#5{%
  \reset@font\fontsize{#1}{#2pt}%
  \fontfamily{#3}\fontseries{#4}\fontshape{#5}%
  \selectfont}%
\fi\endgroup%
\begin{picture}(6809,2499)(1039,-2023)
\put(3676,-736){\makebox(0,0)[lb]{\smash{{\SetFigFont{12}{14.4}{\familydefault}{\mddefault}{\updefault}{\color[rgb]{0,0,0}P}%
}}}}
\put(6601,-1030){\makebox(0,0)[lb]{\smash{{\SetFigFont{12}{14.4}{\familydefault}{\mddefault}{\updefault}{\color[rgb]{0,0,0}$a_{\rm b,1}\left(1\pm e\right)$}%
}}}}
\put(5300,-1030){\makebox(0,0)[lb]{\smash{{\SetFigFont{12}{14.4}{\familydefault}{\mddefault}{\updefault}{\color[rgb]{0,0,0}$a_{\rm b, 2}\left(1\pm e\right)$}%
}}}}
\put(7350,-286){\makebox(0,0)[lb]{\smash{{\SetFigFont{12}{14.4}{\familydefault}{\mddefault}{\updefault}{\color[rgb]{0,0,0}$a-a_{\rm b,1}\left(1\pm e\right)$}%
}}}}
\put(1426,314){\makebox(0,0)[lb]{\smash{{\SetFigFont{12}{14.4}{\familydefault}{\mddefault}{\updefault}{\color[rgb]{0,0,0}Config A}%
}}}}
\put(2776,-211){\makebox(0,0)[lb]{\smash{{\SetFigFont{10}{14.4}{\familydefault}{\mddefault}{\updefault}{\color[rgb]{0,0,0}$r_1 = \sqrt{a_{\rm b,1}^2\left(1\pm e\right)^2 + a^2}$}%
}}}}
\put(1801,-1786){\makebox(0,0)[lb]{\smash{{\SetFigFont{12}{14.4}{\familydefault}{\mddefault}{\updefault}{\color[rgb]{0,0,0}Inner Disk Edge}%
}}}}
\put(2370,-1360){\makebox(0,0)[lb]{\smash{{\SetFigFont{8}{14.4}{\familydefault}{\mddefault}{\updefault}{\color[rgb]{0,0,0}2}%
}}}}
\put(2370,-50){\makebox(0,0)[lb]{\smash{{\SetFigFont{8}{14.4}{\familydefault}{\mddefault}{\updefault}{\color[rgb]{0,0,0}1}%
}}}}
\put(2776,-640){\makebox(0,0)[lb]{\smash{{\SetFigFont{12}{14.4}{\familydefault}{\mddefault}{\updefault}{\color[rgb]{0,0,0}a}%
}}}}
\put(1401,-436){\makebox(0,0)[lb]{\smash{{\SetFigFont{12}{14.4}{\familydefault}{\mddefault}{\updefault}{\color[rgb]{0,0,0}$a_{\rm b,1}\left(1\pm e\right)$}%
}}}}
\put(1401,-1036){\makebox(0,0)[lb]{\smash{{\SetFigFont{12}{14.4}{\familydefault}{\mddefault}{\updefault}{\color[rgb]{0,0,0}$a_{\rm b, 2}\left(1\pm e\right)$}%
}}}}
\put(5851,-1786){\makebox(0,0)[lb]{\smash{{\SetFigFont{12}{14.4}{\familydefault}{\mddefault}{\updefault}{\color[rgb]{0,0,0}Inner Disk Edge}%
}}}}
\put(7726,-736){\makebox(0,0)[lb]{\smash{{\SetFigFont{12}{14.4}{\familydefault}{\mddefault}{\updefault}{\color[rgb]{0,0,0}P}%
}}}}
\put(5820,-770){\makebox(0,0)[lb]{\smash{{\SetFigFont{8}{14.4}{\familydefault}{\mddefault}{\updefault}{\color[rgb]{0,0,0}2}%
}}}}
\put(5251,314){\makebox(0,0)[lb]{\smash{{\SetFigFont{12}{14.4}{\familydefault}{\mddefault}{\updefault}{\color[rgb]{0,0,0}Configs B/C}%
}}}}
\put(7010,-770){\makebox(0,0)[lb]{\smash{{\SetFigFont{8}{14.4}{\familydefault}{\mddefault}{\updefault}{\color[rgb]{0,0,0}1}%
}}}}
\put(2776,-1186){\makebox(0,0)[lb]{\smash{{\SetFigFont{10}{14.4}{\familydefault}{\mddefault}{\updefault}{\color[rgb]{0,0,0}$r_2 = \sqrt{a_{\rm b,2}^2\left(1\pm e\right)^2 + a^2}$}%
}}}}
\end{picture}%
 \caption{Cartoon showing the various geometric configurations of the binary, each leading to a different position of the ice line. The sign of the eccentricity terms depends on whether the stars are at apastron $\left(+\right)$ or periastron $\left(-\right)$. The star labeled as ``1'' is the primary and the star labeled as ``2'' is the secondary. Configurations B and C are geometrically similar, just with the primary and secondary switched. I assume that in configurations B and C, the star nearest to the point P in the circumbinary disk fully eclipses its companion.
  \label{fig:binary_configs}}
\end{figure*}

{\it Irradiation: }Around a binary system the ice line will oscillate due to the orbital motion of the irradiating sources. The extrema bounding the ice line depends on the stellar temperatures $\left(T_{\star, 1}, T_{\star, 2}\right)$, stellar radii $\left(R_{\star, 1}, R_{\star, 2}\right)$, stellar masses $\left(M_{\star, 1}, M_{\star, 2}\right)$, and the ratio of the orbital period of the binary to the cooling time for the disk material $\left(\tau \equiv P_{\rm bin}/t_{\rm cool}\right)$. Figure \ref{fig:binary_configs} shows instantaneous orbital configurations of the binary that produce the two extrema between which the ice line will lie. The semi-major axes of the primary and secondary I define to be $a_{\rm b, 1}$ and $a_{\rm b, 2}$, respectively.

For systems where the cooling time of the disk material is much longer than the binary orbital period ($\tau \ll 1$), the ice line will be static, with a position corresponding to the maximum ice line position of the $\tau \gtrsim 1$ scenario. Since I am interested in the ice line as it relates to planet formation, I only consider the maximum position of the ice line in this study becase in the $\tau \gtrsim 1$ case, planetesimals that form from condensed water-ice within this maximum ice line will be reheated as the binary orbits. Since binary orbital timescales are much shorter than timescales for planet formation, water-ice-based planetesimals within the maximum ice line will be too short lived to form planets.

For configuration A shown in Figure \ref{fig:binary_configs}, the flux incident at a point $P$ in the disk is
\begin{align}
    F_{\rm A} = & {} \left(\frac{\alpha_1}{2}\right)\left(\frac{R_{\star, 1}}{r_1}\right)^2\sigma T_{\star, 1}^4 \nonumber \\
    & {} + \left(\frac{\alpha_2}{2}\right)\left(\frac{R_{\star, 2}}{r_1}\right)^2\sigma T_{\star, 2}^4\; ,
\end{align}
where $r_1=\sqrt{a_{\rm b, 1}^2\left(1\pm e\right)^2+a^2}$, $r_2=\sqrt{a_{\rm b, 2}^2\left(1\pm e\right)^2+a^2}$ and $\alpha$ is the angle at which the irradiation strikes the disk, given by
\begin{equation}
    \alpha \approx \frac{0.4R_{\star}}{a} + \frac{8}{7}\left(\frac{T_{\star}}{8\times10^6~{\rm K}}\right)^{4/7}\left(\frac{a}{R_{\star}}\right)^{2/7}\; . \label{eqn:alpha}
\end{equation}
For the parameters adopted, $\tau_{\rm R} > 1$ for both the incoming radiation and the radiation from the midplane for $a \lesssim 63~$AU. In this region, the inner disk temperature of a passive disk is $T_{\rm irr, A} = T_e/2^{1/4}$, where $T_e = \left(F_{\rm A}/\sigma\right)^{1/4}$, which is independent of disk surface density \citep{1997ApJ...490..368C}.

For configuration B, pictured in Figure \ref{fig:binary_configs}, I assume that the primary fully eclipses the secondary, such that the only flux the disk receives at the point $P$ is from the primary. The flux incident on the disk is
\begin{equation}
    F_{\rm B} = \left(\frac{\alpha_1}{2}\right)\left(\frac{R_{\star, 1}}{a-a_{\rm b, 1}\left(1\pm e\right)}\right)^2\sigma T_{\star, 1}^4\; , \label{eqn:f_b}
\end{equation}
and the temperature of the interior of the disk is $T_{\rm irr,B} = \left(F_{\rm B}/2\sigma\right)^{1/4}$. Configuration C, similar in geometry to B, shown in Figure \ref{fig:binary_configs}, I assume the secondary fully eclipses the primary, yielding an internal disk temperature of $T_{\rm irr,C} = \left(F_{\rm C}/2\sigma\right)^{1/4}$, where $F_{\rm C}$ is given by equation (\ref{eqn:f_b}) but with $R_{\star, 1}\rightarrow R_{\star, 2}$, $T_{\star, 1}\rightarrow T_{\star, 2}$, $\alpha_1\rightarrow\alpha_2$, and $a_{\rm b, 1}\rightarrow a_{\rm b, 2}$. Note that in the case of equal mass binaries, configurations B and C are degenerate.

Combining heating due to accretion and irradiation, the midplane temperature of this circumbinary disk model is given by
\begin{equation}
    T_{\rm mid}^4 = T_{\rm acc}^4 + T_{\rm irr}^4\; . \label{eqn:t_mid}
\end{equation}

\section{Results}
\label{sec:results}
I find the position of the ice line by determining the point in the circumbinary disk where equation (\ref{eqn:t_mid}) is equal to 160~K. Originally, \citet{1981PThPS..70...35H} took the ice line condensation temperature to be 170~K, however, \citet{2004M&PS...39.1859P} showed that the ice line temperature can be as low as $\sim 145~$K. Rather than balancing the evaporative cooling and condensation heating for the grains to find the position of the ice line, I adopt an ice line temperature of 160~K, since the position of the ice line will not change significantly for condensation temperatures in this range.

\begin{figure}[h!]
\epsscale{0.9}
\plotone{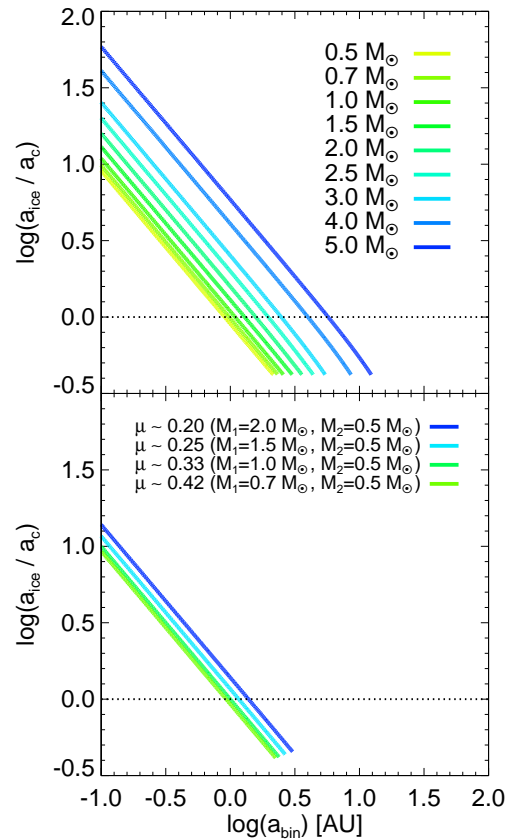}
\caption{Position of the ice line relative to the critical semi-major axis as a function of binary separation for binaries on circular orbits. The top panel shows curves for equal mass binaries ($\mu = 1/2$), with the mass of each component labeled in the legend, while the bottom panel allows for various mass ratios, $\mu \equiv M_2/\left(M_1+M_2\right)$. 
  \label{fig:aice_circ}}
\end{figure}

Figure \ref{fig:aice_circ} shows the maximum ice line position relative to the critical semi-major axis, interior to which orbits are unstable (see equation (\ref{eqn:a_crit})), as a function of the binary separation for both equal mass binaries (top panel) and unequal mass binaries (bottom panel) on circular orbits. The unequal mass systems are constructed from a $0.5~{\rm M_{\odot}}$ star as a secondary. For the $\mu \sim 0.2$, $\mu\sim 0.25$, $\mu\sim 0.33$, and $\mu\sim 0.42$ systems, the primary has a mass of $2~{\rm M_{\odot}}$, $1.5~{\rm M_{\odot}}$, $1~{\rm M_{\odot}}$, and $0.7~{\rm M_{\odot}}$, respectively.

In all systems I consider that have component masses $\lesssim 2.0~{\rm M_{\odot}}$, heating by accretion dominates the determination of the ice line position. Assuming that for separations from the ice line inwards there is only accretional heating, equation (\ref{eqn:t_mid}) can be approximated as $T_{\rm mid} = T_{\rm acc}$. Combining this with equations (\ref{eqn:t_effacc}) and (\ref{eqn:t_acc}), along with the fact that $\tau_{\rm R}\gg 2/3$ in this regime,
\begin{align}
     a_{\rm ice}^{\rm acc} \approx & {} 2.13~{\rm AU}\left(\frac{M_{\rm tot}}{\rm M_{\odot}}\right)^{2/9}\left(\frac{\kappa_{\rm R}}{1~{\rm cm^2~g^{-1}}}\right)^{2/9}\nonumber \\
     & {} \times \left(\frac{\dot{M}}{10^{-8}~{\rm M_{\odot}~yr^{-1}}}\right)^{2/9}\left(\frac{T_{\rm ice}}{160~{\rm K}}\right)^{-8/9} \label{eqn:aice_analytic}
\end{align}
which shows the position of the ice line for a given total binary mass $M_{\rm tot}$ is independent of the binary separation. Normalizing this ice line position to the critical semi-major axis, which is proportional to the binary separation for a given mass ratio and eccentricity, it is clear that $a_{\rm ice}/a_c \propto a_{\rm bin}^{-1}$. This proportionality holds for any binary eccentricity and mass ratio, provided that no component of the binary has a mass $\gtrsim 2.0~{\rm M_{\odot}}$, so that accretion is the dominant heating mechanism out to the position of the ice line.

The most important information to be gleaned from Figure \ref{fig:aice_circ} is that in each case, there is a critical binary separation for which $a_{\rm ice}/a_{\rm c}=1$, which I define as $a_{\rm bin}^{\rm crit} \equiv a_{\rm bin}\left(a_{\rm ice}/a_c=1\right)$. For binary separations equal to and larger than $a_{\rm bin}^{\rm crit}$, the ice line is \emph{always} interior to the inner disk edge. Analytically, employing equation (\ref{eqn:aice_analytic}) and the expression for the critical semi-major axis as a function of binary separation for equal mass binaries on circular orbits ($a_c\approx 2.4a_{\rm bin}$),
\begin{align}
    \frac{a_{\rm ice}^{\rm acc}}{a_c} \approx & {} ~0.89\left(\frac{\rm AU}{a_{\rm bin}}\right)\left(\frac{M_{\rm tot}}{\rm M_{\odot}}\right)^{2/9}\left(\frac{\kappa_{\rm R}}{1~{\rm cm^2~g^{-1}}}\right)^{2/9} \nonumber \\
    & {} ~\times \left(\frac{\dot{M}}{10^{-8}~{\rm M_{\odot}~yr^{-1}}}\right)^{2/9}\left(\frac{T_{\rm ice}}{160~{\rm K}}\right)^{-8/9}\; ,
\end{align}
which is good to within a few percent for all the equal mass systems shown in Figure \ref{fig:aice_circ} with component masses $\lesssim 2.0~{\rm M_{\odot}}$. Setting $a_{\rm ice}^{\rm acc}/a_c = 1$ yields an expression for $a_{\rm bin}^{\rm crit}$.

The equal mass binaries with component masses $\gtrsim 2.5~{\rm M_{\odot}}$ heat their disks primarily by stellar irradiation. Assuming $T_{\rm mid} = T_{\rm irr} = T_{\rm ice}$ for equal mass binaries on circular orbits in configuration A shown in Figure \ref{fig:binary_configs} yields no closed-form solution for $a_{\rm ice}^{\rm irr}/a_c$. However, an analytical expression for $a_{\rm bin}^{\rm crit}$ resulting from heating by stellar irradiation can be obtained:
\begin{align}
    a_{\rm bin}^{\rm crit} \approx & {} 0.198~{\rm AU}\left(\frac{T_{\star}}{4000~{\rm K}}\right)^{8/3}\nonumber \\
    & {} \times \left(\frac{R_{\star}}{\rm R_{\odot}}\right)\left(\frac{T_{\rm ice}}{160~{\rm K}}\right)^{-7/3}\; .
\end{align}

\begin{figure}[h!]
\epsscale{0.9}
\plotone{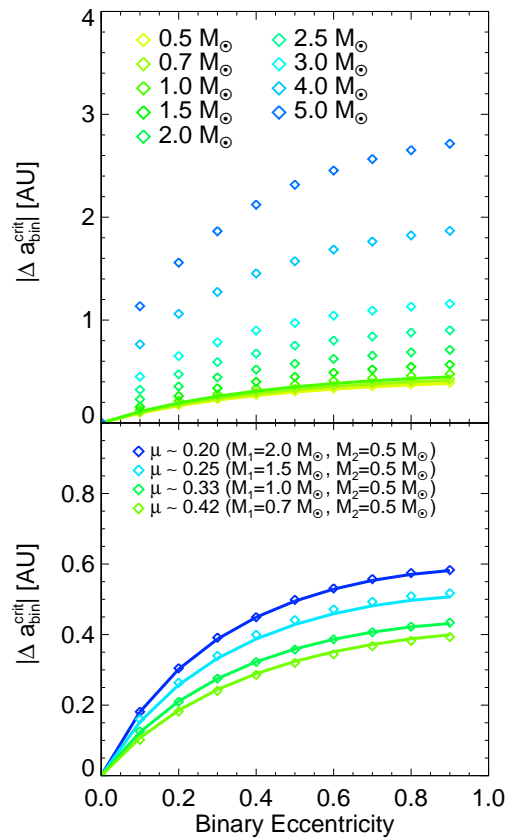}
\caption{Change in the critical binary separation, $a_{\rm bin}^{\rm crit}$, defined to be the binary separation where the ice line is equal to or interior to the critical semi-major axis, $a_{\rm ice} \leq a_c$, relative to circular orbits as a function of binary eccentricity. The top panel shows that for equal mass binaries and the bottom panel shows the same for unequal mass binaries. The diamonds in each plot are calculated points at steps of $\Delta e=0.1$, color coded by component mass or mass ratio. The systems in the accretion-dominated regime have analytical lines drawn according to equation (\ref{eqn:abin_crit_acc_analytic}).
  \label{fig:atr_ecc}}
\end{figure}

If the binaries are allowed to have eccentric orbits, the critical semi-major axis, $a_c$, grows via equation (\ref{eqn:a_crit}). All unequal mass binaries considered and the equal mass binaries with component masses $\lesssim 1.0~{\rm M_{\odot}}$ remain in the accretion-dominated regime for all eccentricities. However, for equal mass binaries with component masses $\gtrsim 1.0~{\rm M_{\odot}}$, irradiation dominates at distances near the ice line for binary separations equal to the critical value, $a_{\rm bin}^{\rm crit}$. In both the accretion- and irradiation-dominated regimes, since $a_c$ always increases, $a_{\rm bin}^{\rm crit}$ decreases with eccentricity. Figure \ref{fig:atr_ecc} shows the change in $a_{\rm bin}^{\rm crit}$ as a function of binary eccentricity for systems of both equal and unequal mass binaries.

For the systems in the accretion-dominated regime, the change in $a_{\rm bin}^{\rm crit}$ as a function of binary eccentricity can be expressed analytically by combining equations (\ref{eqn:aice_analytic}) and (\ref{eqn:a_crit}):
\begin{align}
    \Delta a_{\rm bin}^{\rm crit} \approx & {} \left[\frac{1}{\phi\left(e,\mu\right)} - \frac{1}{\phi\left(0,\mu\right)}\right]2.13~{\rm AU}\left(\frac{M_{\rm tot}}{\rm M_{\odot}}\right)^{2/9}\nonumber \\
    & {} \times \left(\frac{\kappa_{\rm R}}{1~{\rm cm^2~g^{-1}}}\right)^{2/9}\left(\frac{\dot{M}}{10^{-8}~{\rm M_{\odot}~yr^{-1}}}\right)^{2/9}\nonumber \\
    & {} \times \left(\frac{T_{\rm ice}}{160~{\rm K}}\right)^{-8/9}\; , \label{eqn:abin_crit_acc_analytic}
\end{align}
where $\phi\left(e,\mu\right)$ is the coefficient for the critical semi-major axis, given by the expression in the brackets of equation (\ref{eqn:a_crit}). For the accretion-dominated systems in Figure \ref{fig:atr_ecc}, I plot these analytic curves. I find no simple, analytic form for $\Delta a_{\rm bin}^{\rm crit}$ in the irradiation-dominated regime. The main point from Figure \ref{fig:atr_ecc} is that binary eccentricity serves to decrease $a_{\rm bin}^{\rm crit}$ in all cases.

I also examine how the maximum ice line changes in time, as the binary components contract to the main sequence, for equal mass binaries on circular orbits. Using the stellar evolution models of \citet{2000A&A...358..593S}, I calculated $a_{\rm bin}^{\rm crit}$ for each system as a function of time, as the binary components approach the main sequence. For equal mass binaries consisting of stars less massive than $\sim 2~{\rm M_{\odot}}$, the temperature and luminosity do not increase enough to dominate disk heating over steady mass accretion until they reach ZAMS. However, even at ZAMS, the ice line is not increased much from the initial position at formation time. Furthermore, by the time stars less massive than $\sim 2~{\rm M_{\odot}}$ reach the ZAMS, their disks will have dissipated and any planet formation should have already ocurred. Conversely, equal mass binaries consisting of stars more massive than $\sim 2~{\rm M_{\odot}}$ will indeed have their ice line locations significantly increased as they contract towards the main sequence. The effective temperature of such massive stars increases rapidly as they approach the main sequence, and stellar irradiation quickly becomes the dominant heating mechanism for their disks. Since $t_{\rm ZAMS}$ for massive stars is much less than disk lifetimes, any planet formation around massive star binaries will necessarily continue after they reach the main sequence. For the purposes of this study, I thus restrict my conclusions to binaries consisting of low to intermediate mass components, $M_{\star} \lesssim 2~{\rm M_{\odot}}$.

It is also worth noting that the mass accretion rate among T Tauri stars declines as the stars age \citep{1998ApJ...495..385H}. However, since $a_{\rm ice}^{\rm acc} \propto \dot{M}^{2/9}$, a declining mass accretion rate will serve to move the ice line inwards. \citet{2006ApJ...640.1115L} pointed out that the disk lifetime scales as $1/\dot{M}$ and that the lifetime of the minimum-mass solar nebula disk for a constant $\dot{M} = 10^{-8}~{\rm M_{\odot}~yr^{-1}}$ is $2\times10^6~$yr, or roughly the lifetime of the nebular gas. Accretion rates higher than $10^{-8}~{\rm M_{\odot}~yr^{-1}}$ would further constrain the (already difficult to explain) timescales for giant planet formation. Nonetheless, the scaling of the ice line in the accretion-dominated regime with the mass accretion rate is weak and thus variations of an order of magnitude would not significantly change the results of this study.

It is interesting that there is a maximum binary separation (i.e. $a_{\rm bin}^{\rm crit}$) for which systems with separations larger than this maximum value have ice lines that lie interior to the critical semi-major axis. Circumbinary disks around systems with $a_{\rm bin} \geq a_{\rm bin}^{\rm crit}$ will therefore have condensed water-ice throughout the entire disk, meaning that rocky planets should not form in such systems. Using the properties of the observed binary population, I calculate the fraction of these binary systems with $a_{\rm bin} \geq a_{\rm bin}^{\rm crit}$.

\citet{2010ApJS..190....1R} determine the orbital period distribution of solar-type stars with stellar companions to be consistent with a Gaussian, with a mean of $\log{P} = 5.03$ and $\sigma_{\rm \log{P}}=2.28$, where $P$ is in days. I assume that this period distribution found by \citet{2010ApJS..190....1R} is the same for all equal mass binaries with component masses in the range I consider, $0.5~{\rm M_{\odot}\leq M_{\star}\leq 2.0~{\rm M_{\odot}}}$. I then compute the the periods corresponding to $a_{\rm bin}^{\rm crit}$ for equal mass binaries with component masses in this range and integrate over a Gaussian distribution with the above parameters to estimate the fraction of binaries with ice lines interior to their inner disk edge.

For equal mass binaries with total masses $M_{\rm tot} \leq 4.0~{\rm M_{\odot}}$, the percentage of binary systems that have ice lines interior to the critical semi-major axis, at $t_{\rm ZAMS}/10$ and later, ranges from 81\% ($M_{\rm tot} = 4~{\rm M_{\odot}}$) to 84\% ($M_{\rm tot} = 1~{\rm M_{\odot}}$). These fractions represent lower limits because I conservatively choose the ``maximum'' ice line in each system. Furthermore, a fraction of 80\% is a conservative lower limit for unequal mass binaries since $a_{\rm bin}^{\rm crit}$ for the lowest mass ratio system I consider ($\mu\sim 0.2$) is approximately the same as that of the equal mass binary with component masses of $2.0~{\rm M_{\odot}}$.

\section{Conclusion}
\label{sec:conclusion}
I have shown that ice lines lie interior to the critical semi-major axis for $\gtrsim 80\%$ of all equal and unequal mass binaries with components less massive than $M_{\star} \lesssim 2~{\rm M_{\odot}}$. This suggests that rocky planets should not form in these systems; only gas giant and icy planets.

These results can be tested by finding more circumbinary planets, specifically planets around binaries with separations larger than the critical separation. This is possible, albeit difficult, with \emph{Kepler}, because there are fewer eclipsing binaries at larger separations and planetary transit probabilities decrease with orbital period. Microlensing can also detect circumbinary planets at a range of binary separations and seems to be the most promising method for testing the results of this study \citep{2008ApJ...676L..53H}. 

\acknowledgments
I gratefully acknowledge helpful conversations with Todd Thompson and Scott Gaudi about this project, as well as critical readings of the text.

\end{document}